\documentclass[twocolumn,article,amsmath,amssymb,floatfix,aps]{revtex4}
\usepackage{graphicx}
\usepackage{epsfig}
\usepackage{multirow}
\usepackage{color}
\usepackage{cancel}
\usepackage{graphics}
\begin{document}
\title{New results about the canonical transformation for boson operators}

\author{C. M. Raduta $^{a)}$ and A. A. Raduta$^{a), b)}$}

\affiliation{$^{a)}$ Department of Theoretical Physics, Institute of Physics and
  Nuclear Engineering, Bucharest, POBox MG6, Romania}

\affiliation{$^{b)}$Academy of Romanian Scientists, 54 Splaiul Independentei, Bucharest 050094, Romania}
\begin{abstract}
The Bogoliubov transformation for a monopole boson induces an unitary transformation connecting the Fock spaces of initial and correlated boson-s.
Here we provide a very simple method for deriving the analytical expression for the overlap matrix of the basis states generating the two boson spaces.
\end{abstract}
\maketitle
\setcounter{equation}{0}
\renewcommand{\theequation}{1.\arabic{equation}}
\section{Introduction}
The boson operators are widely used in various branches of Physics. In particular, Nuclear Physics benefited of this concept in many respects. Thus, in the many body theories the particle-hole or particle-particle operator are approximated with boson-s which results in having by far  a more tractable method, known under the name of the Random Phase Approximation (RPA)\cite{Kiss}. Going beyond
RPA, one has been introduced the concept of the boson expansion method where pairs of fermion operators are replaced by boson series, afterwards truncated such that their algebra be conserved
\cite{JDFJ,Cea,Klei}. Concerning phenomenological methods, all of them are using in a way or another the boson-s as a device for tackling  various physical problems. To give only a few examples we mention the Interacting Boson Approximation \cite{AI76,ArIa76},  the Coherent State Model \cite{AAR81}, the boson description of the triaxial rotor \cite{Ba79,Ta06,Oi06}. In some of the mentioned cases, for simplicity reasons, the model boson Hamiltonian is truncated at second order. To treat this form, in order to get rid of the cross terms, which raise convergence difficulties, one uses a canonical transformation defining new boson-s in terms of which the dangerous terms do not show up. To each type of boson-s, initial or transformed, one associates a basis generating two boson spaces, respectively. Long time ago, the unitary transformation connecting the mentioned bases was analytically expressed \cite{Ta73,Ras73,Rad77}.

The present letter provides this analytical expression by a distinct method, much simpler than the previous solutions. To touch this goal we organized the paper as follow.
In section 2 we treat a second order boson Hamiltonian H, through the Bogoliubov (B) transformation. In section 3 we use the Bargmann representation, where the eigenvalue equation for H  is transformed
into a Schr\"{o}dinger equation for a harmonic oscillator, whose wave-function belongs to the correlated boson-s basis. Going back to the boson picture the matrix, we are looking for, is readily obtained. A short summary of the procedure is presented in section 4.
\setcounter{equation}{0}
\renewcommand{\theequation}{2.\arabic{equation}}
\section{The boson Hamiltonian}
Here we aim at finding the eigenstates and corresponding eigenvalues of the following boson Hamiltonian:
\begin{equation}
H=\epsilon b^{\dagger}b-X(b^{\dagger 2}+b^2),
\end{equation}
where $b^{\dagger}$ and $b$ are boson operator,i.e. obey the commutation relation:
\begin{equation}
\left[b,b^{\dagger}\right]=1.
\end{equation}
The properties of $H$ are studied within the boson space generated by the basis:
\begin{equation}
|k\rangle =\frac{b^{\dagger k}}{\sqrt{k!}}|0\rangle, k=1,2,3,...
\label{bazab}
\end{equation}
with $|0\rangle $ denoting the vacuum state for the boson operators, obeying the equation $b|0\rangle  =0.$
The boson operator satisfy the equations of motion:
\begin{eqnarray}
&&\left[H,b^{\dagger}\right]=\epsilon b^{\dagger}-2Xb,\nonumber\\
&&\left[H,b\right]=-\epsilon b+2Xb^{\dagger}.
\end{eqnarray}
Since the equations of motion are linear in the boson operators, one can define the linear combination
\begin{equation}
\tilde{b}^{\dagger}=Ub^{\dagger}-Vb,
\label{cantr}
\end{equation}
such that the following equations hold:
\begin{equation}
\left[H,\tilde{b}^{\dagger}\right]=\omega \tilde{b}^{\dagger},\;\;\;
\left[\tilde{b},\tilde{b}^{\dagger}\right]=1.
\label{harm}
\end{equation}
These equations assert that the new operators are also of boson type and moreover, in terms of the new operators the Hamiltonian is harmonic.
The first equation (\ref{harm}) provides a homogeneous system of equations, for the amplitudes $U$ and $V$, whose compatibility condition leads to the following expression for the
energy $\omega$:
\begin{equation}
\omega =\epsilon\sqrt{1-4\kappa^2},\;\;\rm{with}\;\; \kappa=\frac{X}{\epsilon}.
\end{equation}
Noticeable the fact that the defined excitation energy $\omega$ exists if $|\kappa|\le 1/2$. For $\epsilon=\pm 2X$ the solution is vanishing which results a critical value for a phase 
transition \cite{Gol61}.
The system becomes unstable for $X> \frac{1}{2}\epsilon$ or $X< -\frac{1}{2}\epsilon$
As for the unknowns $U$ and $V$, they are determined up to a multiplicative constant to be fixed by the normalization equation given by the second equation (\ref{harm}):
\begin{equation}
U^2-V^2=1.
\end{equation}
The result for $U$ and $V$ is :
\begin{eqnarray}
\left(\begin{matrix}U^2\cr V^2\end{matrix}\right)=\frac{1}{2}\left(\pm1+\frac{1}{\sqrt{1-4\kappa^2}}\right).
\end{eqnarray}
To the newly defined boson operators, one associates the basis states:
\begin{equation}
\tilde{|m\rangle}=\frac{\tilde{b}^{\dagger m}}{\sqrt{m!}}\tilde{|0\rangle},\;\;\tilde{b}\tilde{|0\rangle}=0.
\end{equation}
The goal of the present paper is to provide an analytical expression for the overlap matrix:
\begin{equation}
G_{mn}=\langle m|\tilde{n}\rangle.
\end{equation}
This objective will be accomplished in the next section.
Note that the transformation (\ref{cantr}) can be written in an alternative form:
\begin{eqnarray}
\tilde{b}^{\dagger} &=& Tb^{\dagger}T^{\dagger}= Ub^{\dagger}-Vb,\nonumber\\
\tilde{b}           &=& TbT^{\dagger}= Ub-Vb^{\dagger},\;\;\rm{with}\nonumber\\
T&=&e^S,\;\;S=\frac{1}{2}y(b^{\dagger 2}-b^2),\;\;\rm{and}\nonumber\\
U&=&\cosh y,\;\; V=\sinh y.
\label{canT}
\end{eqnarray}
The transformation (\ref{cantr}) can be reversed, and thus $H$ can be expressed in terms of the new boson operators. The result is:

\begin{equation}
H=-\frac{1}{2}(\epsilon-\omega)+\omega \tilde{b}^{\dagger}\tilde{b} .
\end{equation}
Note that in terms of the new boson-s the cross term $\tilde{b}^{\dagger 2}+b^{2}$ does not show up.
The transformation (\ref{cantr}) is known under the name of the Bogoliubov transformation.
\setcounter{equation}{0}
\renewcommand{\theequation}{3.\arabic{equation}}
\section{An alternative treatment of $H$}
For what follows it is useful to introduce the Bargmann representation for the boson $b^{\dagger}$ and $b$, through the mapping
\begin{equation}
b^{\dagger}\to x,\;\;b\to \frac{d}{dx},
\end{equation}
where $x$ denotes a real variable.
Thus, the eigenvalue equation associated to $H$ becomes a differential equation:
\begin{equation}
\epsilon x\frac{d\Psi}{dx}-X\frac{d^2\Psi}{dx^2}-Xx^2\Psi=E\Psi .
\label{difeq}
\end{equation}
By a suitable change of function:
\begin{equation}
\Psi=e^{\frac{x^2}{4\kappa}}\Phi ,
\end{equation}
one gets rid of the first order derivative term and (\ref{difeq}) acquires the Schr\"{o}dinger form:
\begin{equation}
-\frac{1}{2m}\frac{d^2\Phi}{dx^2}+\frac{m\omega^2}{2}x^2\Phi=(E+\frac{\epsilon}{2})\Phi,
\label{Schr}
\end{equation}
where
\begin{equation}
m=\frac{1}{2V},
\end{equation}
and the unit system of $\hbar =c=1$ has been used.
We recognize in Eq. (\ref{Schr}), the Schr\"{o}dinger equation for a linear oscillator of mass $m$ and frequency $\omega$. The oscillator energies are:
\begin{equation}
E_n+\frac{\epsilon}{2}=\omega(n+\frac{1}{2}), n=0,1,2,....
\end{equation}
By comparison, one finds out that these are just the eigenvalues of $H$, derived in the previous section. Correspondingly, the eigenfunction $|\tilde{n}\rangle$ coincides with the function:
\begin{eqnarray}
\Psi_n&=&C_nH_n(\frac{x}{r})e^{\left(-\frac{1}{2}m\omega+\frac{1}{4\kappa}\right)x ^2}\nonumber\\
      &=&C_nH_n(\frac{x}{r})e^{V^2\frac{x^2}{r^2}}.
\end{eqnarray}
Here $H_n$ denotes the Hermite polynomial of rank $n$, $C_n$ is the normalization factor, while $r$  stands for the oscillator length defined by:
\begin{equation}
r^2=\frac{1}{m\omega}=2UV.
\end{equation}
One can check that:
\begin{equation}
-\frac{1}{2}m\omega+\frac{1}{4\kappa}=\frac{V^2}{r^2}.
\end{equation}
Using the analytical expression for the Hermite polynomial and the Taylor expansion for the exponential function one obtains:
\begin{equation}
|\tilde{n}\rangle =C_n\sum_{m,p}\frac{(-1)^{\frac{n-p}{2}}n!2^pV^{m-p}(\frac{x}{r})^{m}}{(\frac{n-p}{2})!p!(\frac{m-p}{2})!}.
\label{Cn}
\end{equation}
The constant $C_n$ can be determined either by brute calculations, restricting the norm of $|\tilde{n}\rangle $ be equal to unity or by the more elegant procedure described in Appendix A.
Using the result from there, (\ref{T0n}), and the obvious relation
\begin{equation}
\langle 0|\tilde{n}\rangle =\frac{(-1)^{\frac{n}{2}}}{\frac{n}{2}!}C_n=\frac{T_{0,n}}{\sqrt{n!}},
\end{equation}
one arrives at:
\begin{equation}
C_n=\frac{V^{\frac{n}{2}}U^{-\frac{m+1}{2}}}{2^{\frac{n}{2}}\sqrt{n!}}.
\end{equation}
We recall that we use the Bargmann representation and therefore the vacuum state normalized to unity is $|0\rangle =1$ and $|m\rangle =\frac{b^{\dagger m}}{\sqrt{m!}}|0\rangle$. Multiplying, to the left, the equation (\ref{Cn}) with $|m\rangle$, and replacing $C_n$ with the expression just obtained, we get:
\begin{equation}
\langle m|\tilde{n}\rangle =\sqrt{m!n!}U^{-\frac{m+n+1}{2}}\sum_{p}\frac{(-1)^{\frac{m-p}{2}}\left(\frac{V}{2}\right) ^{\frac{m+n}{2}-p}}{\left(\frac{m-p}{2}\right)!\left(\frac{n-p}{2}\right)!p!}.
\label{finG}
\end{equation}
Obviously, the summation index $p$ is subject to the restrictions:
\begin{equation}
m-p=even,\;\;n-p=even,\;\;p\le\min\{m,n\}.
\end{equation}
Moreover, $|m-n|=even$.

This expression (\ref{finG}) is identical to those obtained in Refs.\cite{Ta73,Ras73,Rad77} by more tedious methods.

\section{Summary}

In the present paper we derived analytical expression for eigenvalues and eigenstates of a second order boson Hamiltonian by using two alternative methods: a)through a canonical transformation
the Hamiltonian is brought to a diagonal form. The eigenstates are correlated multiphonon states. We looked for analytical expression for the overlap matrix of correlated and non-correlated bosons.
b) In the Bargmann representation the eigenvalue relation associated to H becomes the Schr\"{o}dinger equation for a harmonic oscillator.The wave-function, which is a product of an exponential function and a Hermite polynomial, provides the searched overlap matrix as coefficients  of the associated Taylor series. The result is identical with that previously obtained by one of us (A.A.R.)
by a different likewise more tedious method. Concluding the main result consists in the expression (\ref{finG}) describing the mentioned overlap matrix. As mentioned already, this expression is very useful in many formalisms dealing with monopole bosons.

\setcounter{equation}{0}
\renewcommand{\theequation}{A.\arabic{equation}}
\section{Appendix A}
Here we shall derive the analytical expression for the constant $C_n$ involved in Eq. (\ref{Cn}).

Let us denote by $T_{m,n}$ the matrix elements of the canonical transformation (\ref{canT}) in the basis (\ref{bazab}):
\begin{equation}
T_{m,n}(y)=\langle 0|b^mTb^{\dagger n}|0\rangle .
\end{equation}
From here, one easily obtains the iterative equations:
\begin{eqnarray}
T_{0,n}&=&-VT_{1,n-1},\nonumber\\
T_{1,n-1}&=&U(n-1)T_{0,n-2}-VT_{0,n}.
\end{eqnarray}
which leads to:
\begin{equation}
T_{0,n}=-\frac{V}{U}(n-1)T_{0,n-2} .
\end{equation}
Applying successively this iterative relation, one finds:
\begin{equation}
T_{0,n}=(-\frac{V}{U})^{\frac{n}{2}}\frac{n!}{2^{\frac{n}{2}}(\frac{n}{2})!} T_{0,0}.
\end{equation}
As for $T_{0,0}$ it satisfies the differential equation
\begin{equation}
\frac{d}{dy}T_{0,0}=T_{0,2}=-\frac{V}{2U}T_{0,0}.
\end{equation}
and the initial condition:
\begin{equation}
T_{0,0}(0)=1.
\end{equation}
The solution is:
\begin{equation}
T_{0,0}=U^{-1/2}.
\end{equation}
Consequently:
\begin{equation}
T_{0,n}=(-V)^{\frac{n}{2}}U^{-\frac{n+1}{2}}\frac{n!}{2^{\frac{n}{2}}(\frac{n}{2})!} .
\label{T0n}
\end{equation}

\end{document}